\documentclass[sigconf,nonacm]{acmart}
\AtBeginDocument{%
  }

\setcopyright{acmlicensed}
\copyrightyear{2018}
\acmYear{2018}
\acmDOI{XXXXXXX.XXXXXXX}
\acmConference[Conference acronym 'XX]{Make sure to enter the correct
  conference title from your rights confirmation email}{June 03--05,
  2018}{Woodstock, NY}
\acmISBN{978-1-4503-XXXX-X/2018/06}

\usepackage{xspace}

\newcommand{\sys}{\textsc{Cortex}\xspace}

\begin{document}

\title{Cortex: Workflow-Aware Resource Pooling and Scheduling for Agentic Serving}

\author{Nikos Pagonas}
\email{n.pagonas@columbia.edu}
\affiliation{%
  \institution{Columbia University}
  \city{}
  \country{}
}

\author{Yeounoh Chung}
\email{yeounoh@google.com}
\affiliation{%
  \institution{Google}
  \city{}
  \country{}
}

\author{Kostis Kaffes}
\email{kkaffes@cs.columbia.edu}
\affiliation{%
  \institution{Columbia University}
  \city{}
  \country{}
}

\author{Arvind Krishnamurthy}
\email{arvindkrish@google.com}
\affiliation{%
  \institution{Google \& Univ. of Washington}
  \city{}
  \country{}
}


\begin{abstract}
We introduce \sys, a prototype workflow-aware serving platform designed for agentic workloads.
The core principle of \sys is stage isolation: it provisions dedicated resource pools for each distinct stage of an agentic workflow.
This simple yet powerful strategy mitigates inter-stage interference in compute and memory, leading to better KV cache utilization, higher throughput, and more predictable performance.
By customizing resource allocation and scheduling within each distinct stage of agentic workflows, \sys lays the groundwork for more advanced, agent-native serving paradigms, including malleable resource management, speculative execution of workflow branches, and a shared, multi-tiered cache for ``agentic state.''

\end{abstract}

\maketitle

\section{Motivating Example: Serving Agentic NL2SQL Workflows}
\label{sec:example}

Agentic workflows pair an LLM’s reasoning with iterative tool use: the model observes an intermediate result, thinks, calls another tool, and repeats until the task is solved or a budget is exhausted.
This closed-loop recipe underlies our running example, a Natural-language-to-SQL (NL2SQL) agent that turns a plain-English question like ``What were Europe's sales last quarter?'' into a successfully executed SQL query.
A production-grade NL2SQL workflow typically (i) retrieves the target schema, (ii) autoregressively generates a candidate query, (iii) executes it, (iv) verifies the result set, and, if the query fails, (v) fixes it and retries up to a policy-defined limit~\cite{chung2025long,chang2024nl2sql}.
Each turn may terminate in one of three ways: the query executes correctly, it raises a syntax error, or it returns an empty result that signals a semantic failure.
The whole pipeline repeats until the agent either satisfies the incoming NL2SQL request or exhausts its retry budget.
As shown in Figure~\ref{fig:cortex-overview},
in our example workflow, we omit the schema retrieval and verification stages for simplicity and clarity.

Because every stage feeds directly into the next, the end-to-end latency of this NL2SQL loop depends less on any single LLM call than on how well the serving stack coordinates the entire chain.
Yet today’s LLM platforms remain workflow-agnostic.
For example, popular LLM serving frameworks~\cite{vLLM,zheng2024sglang} treat every stage as an isolated LLM call and schedule them in FCFS order, while LLM agent serving platforms, like Autellix~\cite{Autellix}, use more sophisticated prioritization to improve the throughput of the agentic AI workloads, but without awareness of the internal workflow structure. 
HEXGEN-TEXT2SQL~\cite{peng2025hexgen} schedules NL2SQL agentic AI workflow requests based on their remaining deadline slack and estimated execution time across a pool of  LLM serving instances of varying capabilities, blind to internal workflow stages.
We argue that these platforms leave a lot of performance on the table.
They miss obvious caching opportunities; for instance, five refinement attempts against the same schema incur five identical prompt builds and five identical warm-cache SQL executions.
They schedule LLM calls without knowledge of the remaining workflow, oblivious to downstream costs (e.g., a slow SQL back-end), and thus cannot accelerate stragglers that gate the end-to-end latency.
As a result, operators either over-provision expensive general-purpose instances or suffer frequent SLO violations.

\textbf{Goal:} We propose a new workflow-aware serving platform that starts from a call graph (tool calls, LLM calls) and meets user-defined SLOs by provisioning just enough resources and prioritizing requests and calls accordingly.

\section{\sys: Dedicated Engine Pools Per Workflow Stage} 
\label{sec:cortex}

\begin{figure}
    \centering
    \includegraphics[width=\linewidth]{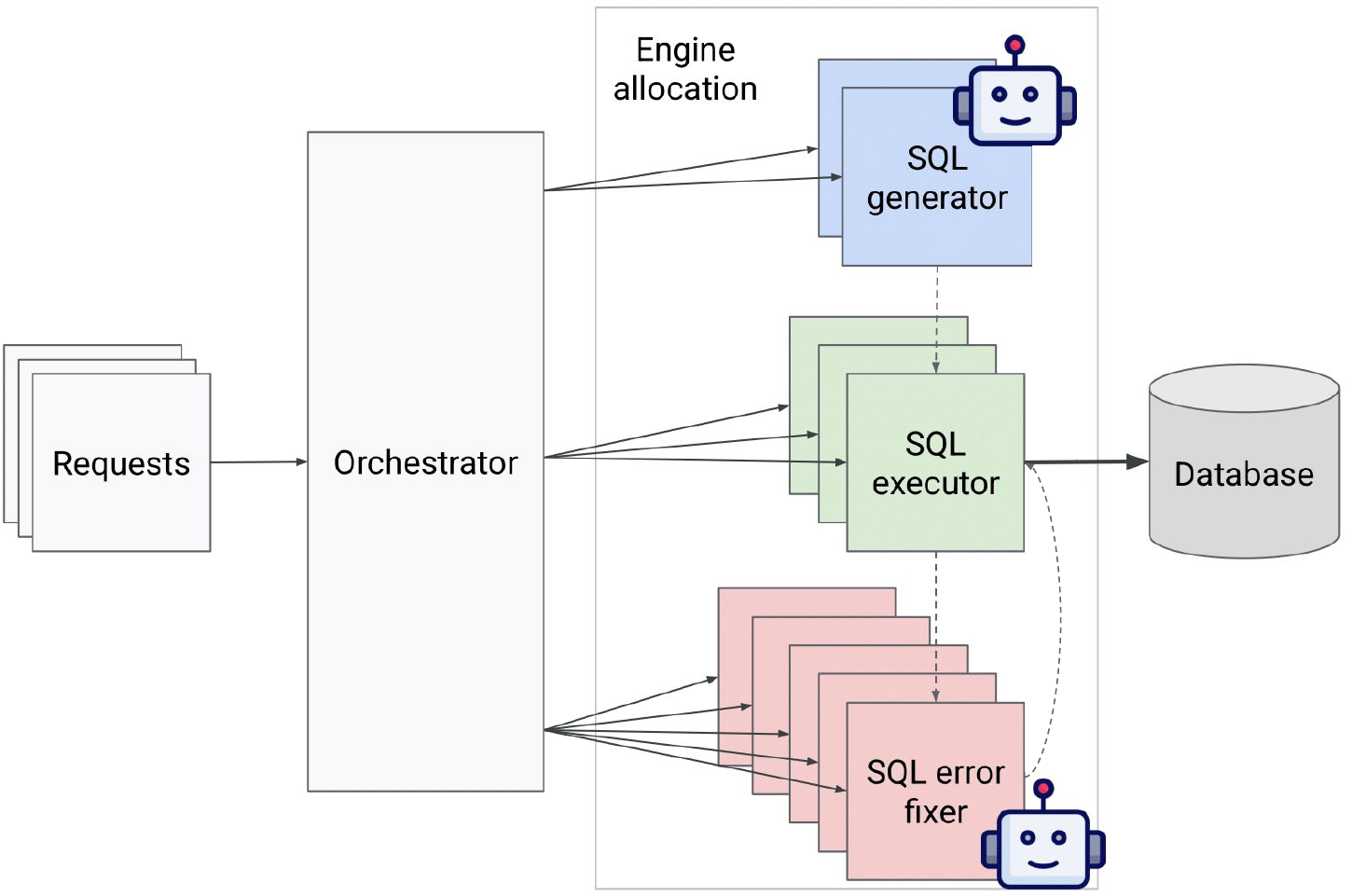}
    \caption{NL2SQL workflow mapped to the \sys architecture. The SQL generator and SQL error fixer stages are LLM calls, while the SQL executor runs candidate SQL queries on a database. \sys provisions separate and dedicated resource pools for each stage.} 
    \label{fig:cortex-overview}
\end{figure}

\begin{figure}
    \centering
    \includegraphics[width=\linewidth,trim=0 40 0 0,clip]{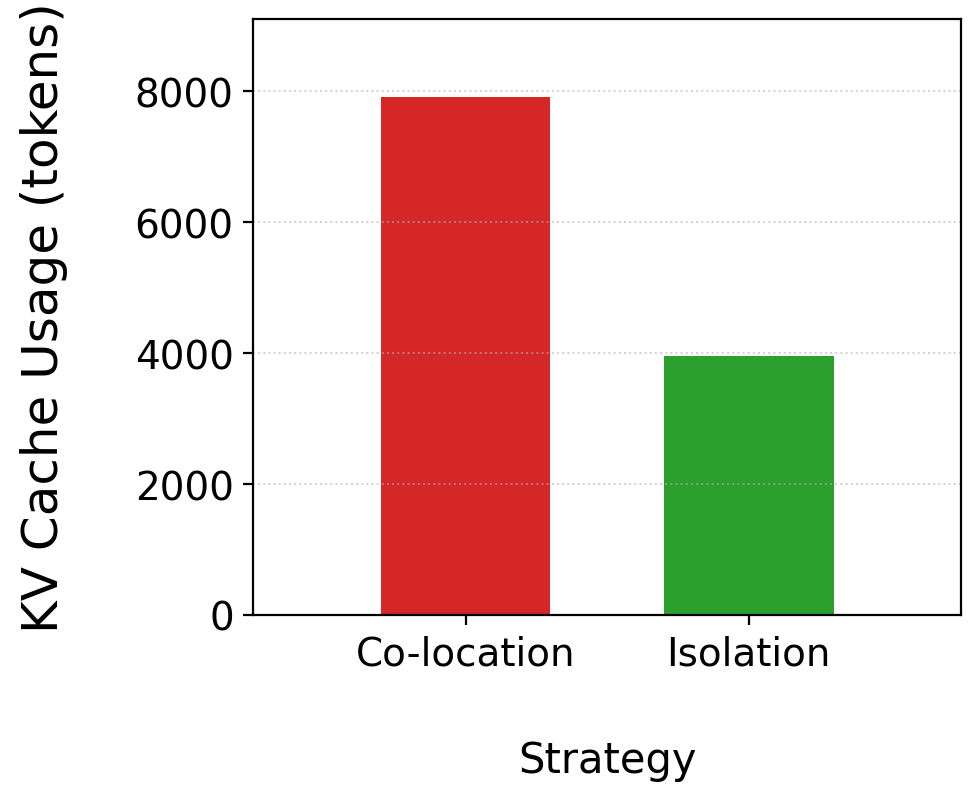}
    \caption{KV Cache usage of the NL2SQL workflow's LLM stages (SQL generator, SQL error fixer) when running in \sys. The total usage is significantly lower when each stage runs in an isolated serving engine.}
    \label{fig:kv-cache-usage}
\end{figure}

In this section, we present \sys, the first prototype of such a platform.
The guiding insight behind \sys is that a single shared pool of ``generic'' LLM engines is a poor fit for agentic workflows whose call graphs contain heterogeneous stages.
Each stage (SQL generation, execution, error fixing) has distinct latency profiles, memory requirements, and caching opportunities. 
Worse, the graph itself can evolve at runtime: the number of parallel explorations, the depth of refinement loops, and the cost of each operator fluctuate with the input and with queueing delays.

\sys addresses this by provisioning a \emph{dedicated engine pool for every workflow stage}.
An engine pool is a homogeneous set of workers—e.g., GPUs for LLM decoding or CPU executors for SQL—managed by a stage-local scheduler with its own queue, cache, and scaling policy. 
Figure~\ref{fig:cortex-overview} sketches the architecture: client requests enter an orchestrator that consumes a compiled call graph, computes per-request SLO slack, and dispatches work to stage pools that are then managed by an engine-allocation layer.

The orchestrator is workflow-aware: it tracks each request’s position in the graph, predicts the next set of eligible operators, and attaches a priority key derived from SLO slack, stage selectivity, and expected service time.
The engine-allocation layer then routes the sub-call to a concrete pool instance that maximizes locality (e.g., KV-cache/prompt affinity for generation; warm connections for SQL), balances load across replicas, reorders requests based on priorities, and enforces admission control when a stage becomes the bottleneck.
When both load and memory pressure are sufficiently low, the orchestrator can opportunistically let compatible stages borrow idling engines to reduce fragmentation and raise utilization.

\noindent{\textbf{\sys Benefits.}}
In this presentation, we demonstrate a straightforward but significant benefit of the stage isolation that is core to \sys's design.
Because each stage carries a distinct prompt and exemplar set, stage isolation markedly improves KV-cache utilization.
As shown in Figure~\ref{fig:kv-cache-usage}, the total KV footprint of the NL2SQL LLM stages (SQL generator and SQL error fixer) is significantly lower when each runs in its own \sys\ pool: each engine holds only its stage-specific context.
In contrast, a shared engine must keep \emph{both} stages’ contexts hot on \emph{each} replica, effectively duplicating KV-cache memory usage.
The reclaimed GPU memory raises effective batch size (and/or beam width) and translates directly into higher throughput and tighter tail latency.

Cortex’s stage isolation yields additional benefits.
First, it eliminates cross-stage interference that wrecks predictability: when heterogeneous calls (e.g., SQL generation and error fixing) share an engine, batching couples their runtimes, delays token emission, and makes an LLM call’s latency depend on its batch-mates—undermining remaining-time/SLO-aware prioritization.
Isolating stages restores stable, stage-local latency models. 
Second, it enables independent scaling and provisioning: a fast monitor scales out or in only the pool that threatens the SLO, letting us lightly provision run-once stages like SQL generation while weighting critical-path pools for the SQL error fixer more heavily.
In NL2SQL, where per-stage variance is far smaller than end-to-end variance (dominated by variable retry depth), this per-stage control can meet the same SLOs with much higher efficiency.

\section{Toward Agent-Native Serving}
\label{sec:discussion}
Stage isolation is only a starting point, hinting at broader still-unrealized mechanisms needed for native agentic serving.

\textbf{Malleable workflows \& resources.}
Because agentic workflows are malleable in both \emph{what} they compute and \emph{how} they run, our platform can flex along two axes.
First, the computation can shrink to fit scarce resources: when latency approaches an SLO bound or cluster load spikes, the planner can swap a heavyweight model for a lighter variant, prune retries, or shorten reasoning, lowering memory pressure and generation time.
Second, resources can stretch to match a demanding computation: in fan-out patterns such as NL2SQL, where dozens of candidate queries execute in parallel, the scheduler can boost the straggling tail using beefier engines with more GPUs  or higher SQL concurrency, equalizing completion times and tightening SLO variance.
The two strategies interplay: if a small-model engine is saturated yet a large-model engine sits idle waiting for refinement, the system can opportunistically let the larger model generate candidates itself, provided it can access or reconstruct the necessary context. 
A central open challenge is designing an interface that lets workflows declare their malleability, i.e., what knobs (model size, parallelism, retry depth) can be tuned and at what cost, and exposing that information uniformly to both the planner and the scheduler.

\textbf{Speculation.}
With their probabilistic reasoning and many viable next actions, LLM agents are inherently fertile ground for speculation.
For example, \sys can speculate on the most probable branches taken in a workflow and, if possible, pre-warm the associated engines or even pre-execute the next steps.
Another possibility is hedging.
For example, in the NL2SQL workflow, the platform can generate many candidate queries and evaluate them in parallel instead of serially.
Speculating on the next actions taken, and preparing them or even pre-executing them, will deliver faster end-to-end responses without waiting for the agent’s final, deterministic choice.

\textbf{Agentic State.}
\sys can eventually treat intermediate data, from KV-cache entries to fetched tool results, as a first-class, multi-tier ``agentic state''.
The simplest form of caching is already implemented by SGLang: each engine keeps its own prompts, schema embeddings, and partial plans resident in GPU memory, preventing thrashing when successive calls share the same context.
Above that, a workflow-wide shared tier could act as a publish/subscribe fabric: agents can advertise artifacts such as previously executed SQL results or vector-search hits, and downstream calls check this tier before recomputing or refetching.
This shared tier would turn repeated tool and LLM calls across concurrent agents into zero-cost hits and sharply reduce both latency and compute waste.


\bibliographystyle{ACM-Reference-Format}
\bibliography{bib}

\end{document}